\documentclass[%
 reprint,
superscriptaddress,
 amsmath,amssymb,
 aps,
]{revtex4-2}

\usepackage{graphicx}
\usepackage{dcolumn}
\usepackage{bm}

\usepackage{adjustbox}
\usepackage{hyperref}
\usepackage{xcolor}
\usepackage{float}
\begin{document}

\title{Different critical exponents on two sides of the magnetic transition in two dimensions}

\author{P. Biswal}
\email{Contact author: pujalin.biswal@iopb.res.in}
\affiliation{Institute of Physics, Bhubaneswar 751005, India}
\affiliation{Homi Bhabha National Institute, Anushaktinagar, Mumbai 400094, India}
\author{Ruma Khatun}
\affiliation{Institute of Physics, Bhubaneswar 751005, India}
\affiliation{Homi Bhabha National Institute, Anushaktinagar, Mumbai 400094, India}
\author{G. Tripathy}
\affiliation{Institute of Physics, Bhubaneswar 751005, India}
\affiliation{Homi Bhabha National Institute, Anushaktinagar, Mumbai 400094, India}
\author{Diptikanta Swain}
\affiliation{Institute of Chemical Technology–IndianOil Odisha Campus, Bhubaneswar 751013, India}
\author{D. Samal}
\email{Contact author: dsamal@iopb.res.in}
\affiliation{Institute of Physics, Bhubaneswar 751005, India}
\affiliation{Homi Bhabha National Institute, Anushaktinagar, Mumbai 400094, India}

\begin{abstract}
The possibility of different critical exponents on two sides of a magnetic transition has been theoretically predicted [PRB 13, 2222 (1976), PRL 115, 200601 (2015)] and experimentally realized [PRL 128, 015703 (2022)] in three-dimensional systems. However, the emergence of such a scenario in two dimensions is not yet established. Here, we report distinct two-dimensional (2D) critical exponents below and above the magnetic transition temperature (T$_C$) through critical scaling analysis in a quasi-2D non-centrosymmetric, Dzyaloshinskii–Moriya interaction (DMI) active hybrid perovskite (C$_7$H$_9$NBr)$_2$CuBr$_4$. The system exhibits critical exponents of 2D Ising-type below T$_C$ and 2D XY-type above T$_C$. In contrast, isostructural Cl and centrosymmetric Br analogs, in which the DMI effect is expected to be weaker, exhibit symmetric 2D critical behavior. These results highlight the crucial role of DMI in (C$_7$H$_9$NBr)$_2$CuBr$_4$ as a symmetry-breaking perturbation that alters the magnetic critical behavior differently above and below T$_C$. Our study is the experimental realization of hitherto unreported asymmetric 2D critical behavior in structurally flexible and dimensionally tunable hybrid perovskites.

\end{abstract}

\maketitle

\begin{figure}
    \centering
    \includegraphics[scale=0.73]{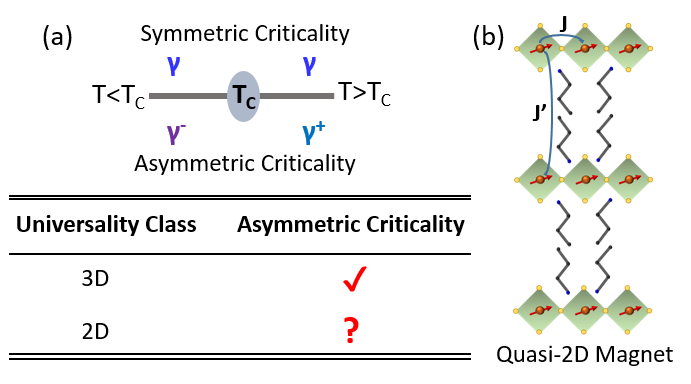}
    \caption{(a) Upper panel represents symmetric (identical $\gamma$) and asymmetric ($\gamma_+\ne\gamma_-$) criticality, Table illustrates the theoretical and experimental realization of asymmetric criticality in 3D systems (Ref. No.:~\cite{PhysRevLett.115.200601,PhysRevLett.128.015703,PhysRevB.13.2222,PhysRevB.107.144425,y5tx-5vm7,10.1063/5.0300224}), which remains elusive in 2D. (b) Schematic of OIHP based quasi-2D magnet.}
    \label{TOC}
\end{figure}

 A second-order magnetic phase transition is characterized by a unique set of critical exponents $\beta$, $\gamma$, and $\delta$, which govern the scaling behavior of the order parameter, susceptibility, and critical isotherm near the transition temperature T$_C$ respectively~\cite{Fisher1967,PhysRevB.96.054406,PhysRevB.79.214426}. Within the renormalization group framework, the critical exponents are associated with the symmetry of order parameter (n) and lattice dimensionality (d) of the system, through their correspondence to the established universality classes (UCs) including mean-field ($\beta$=0.5, $\gamma$=1, and $\delta$=3) three-dimensional (3D) Heisenberg ($\beta$=0.365, $\gamma$=1.386, and $\delta$=4.8), 3D XY($\beta$=0.345, $\gamma$=1.316, and $\delta$=4.81), 3D Ising($\beta$=0.325, $\gamma$=1.24, and $\delta$=4.82), tricritical mean field ($\beta$=0.25, $\gamma$=1, and $\delta$=5), and two-dimensional (2D) Ising ($\beta$=0.125, $\gamma$=1.75, and $\delta$=15) models~\cite{PhysRevB.96.134410,PhysRevB.101.214413,PhysRevB.103.144432,PhysRevB.95.245212,PhysRevB.104.094405,PhysRevMaterials.8.114418,PhysRevB.110.L060405,PhysRevResearch.2.013013}. Conventionally, a single set of critical exponents corresponding to a specific UC is expected to capture the critical behavior across the phase transition region. Notably, Nelson way back in 1976 proposed that symmetry-breaking perturbations such as cubic and hexagonal anisotropy in a 3D O(2) model can induce asymmetric critical behavior, with distinct exponents $\gamma_+$ and $\gamma_-$, above and below T$_C$ respectively~\cite{PhysRevB.13.2222}. He attributed this asymmetry to the dangerously irrelevant nature of the anisotropy, which becomes asymptotically irrelevant at the critical fixed point, but can couple differently in the two sides of transition, thereby modifying the scaling behavior. Nearly five decades later, Leonard and Delamotte proposed a generic mechanism that relies on the possibility of explicitly breaking a continuous symmetry down to discrete one through perturbations that are irrelevant at the critical fixed point, but can generate different critical exponents on the two sides of a continuous phase transition ~\cite{PhysRevLett.115.200601}. Yet, experimental realization of the above striking theoretical proposal remains sparse. The first experimental evidence of asymmetric critical behavior was reported in the skyrmion host magnet  Cu$_2$OSeO$_3$, with the asymmetry linked to the subtle influence of underlying Dzyaloshinskii–Moriya interaction (DMI)~\cite{PhysRevLett.128.015703}. Recent studies have also reported similar unconventional critical behavior in Cr$_{\frac{1}{3}}$TaS$_2$~\cite{PhysRevB.107.144425}, MnBi$_4$Te$_7$ ~\cite{y5tx-5vm7} and FePd$_2$Te$_2$~\cite{10.1063/5.0300224}. However, these experimental and theoretical realizations have thus far been limited to 3D magnetic systems, while the possibility of asymmetric scaling in 2D remains elusive [FIG.~\ref{TOC}(a)].

The 2D limit with the enhanced fluctuations, provides a sensitive regime in which weak symmetry-breaking perturbation can strongly influence the critical behavior and the resulting magnetic order. For an isotropic 2D Heisenberg system, Mermin-Wagner theorem forbids spontaneous symmetry breaking at finite temperature~\cite{PhysRevLett.17.1133}. However, magnetic anisotropy can break the continuous symmetry, suppress fluctuations, and thereby stabilize long-range magnetic order (LRMO), making it central to the study of atomically thin magnets~\cite{Gibertini2019,Burch2018,https://doi.org/10.1002/adma.201900065}. Strong uniaxial anisotropy has been found to drive the system towards the 2D Ising UC in atomically thin magnets such as Cr$_2$Ge$_2$Te$_6$~\cite{Gong2017}, CrI$_3$~\cite{Huang2017}, Fe$_3$GeTe$_2$\cite{Fei2018,Xiao2026}. By contrast, easy-plane anisotropy leads to Berezinskii–Kosterlitz–Thouless (BKT) physics, where vortex–antivortex binding produces quasi-long-range order~\cite{doi:10.1126/science.abd5146,berezinskii1972destruction,kosterlitz1973ordering}. Recently, Parkin \textit{et. al} demonstrated robust 2D XY-type long-range magnetic order ($\beta$$\sim$0.22) in monolayers of CrCl$_3$~\cite{doi:10.1126/science.abd5146}. Between these two extreme cases, systems having fourfold symmetry belong to 2D XYh$_4$ UC with critical exponents lying between 2D Ising and XY limits~\cite{doi:10.1126/science.1221878,Taroni_2008}. These results reveal that symmetry-breaking perturbations such as anisotropy play a decisive role in determining criticality in two dimensions. Whether symmetry-breaking perturbations can drive asymmetric critical scaling in 2D magnetic systems still remains an open question.  Leonard and Delamotte pointed out that it would be extremely interesting  to investigate the possibility for reliazing  asymmetric criticality in 2D~\cite{PhysRevLett.115.200601}. Addressing this question experimentally requires a system that remains predominantly 2D while retaining the specific symmetry-breaking interaction which could generate different critical exponents below and above T$_C$. In particular, quasi-2D organic–inorganic hybrid perovskites (OIHPs) provide a natural platform for realizing 2D magnetic regime  since the interlayer distance in these systems can be tuned to a large value that suppresses the interlayer exchange J’, with J'/J $\sim$ 10$^{-4}$-10$^{-6}$ [FIG.~\ref{TOC}(b)], such that the intralayer exchange J  dominates the magnetic response~\cite{6rgd-57pc,PhysRevMaterials.8.024409,https://doi.org/10.1002/adfm.202207988,SEPTIANY2022168598}. Besides, the structural flexibility of OIHPs enables the emergence of symmetry-breaking interactions such as DMI through broken inversion symmetry induced by the appropriate choice of organic cations.

Here, we demonstrate asymmetric criticality in a non-centrosymmetric, DMI active quasi-2D hybrid perovskite (C$_7$H$_9$NBr)$_2$CuBr$_4$. Remarkably, the system exhibits 2D Ising-type criticality below T$_C$ and 2D XY-type criticality above T$_C$, providing the first experimental realization of asymmetric 2D criticality. To further substantiate our claim, we have also investigated the critical behavior of an isostructural Cl-analog (C$_7$H$_9$NBr)$_2$CuCl$_4$ and centrosymmetric Br-analog (C$_7$H$_{10}$N)$_2$CuBr$_4$, where the DMI effect is expected to be weaker, and we observe a conventional symmetric critical behavior. We attribute the asymmetric criticality in NCS-Br to the symmetry-breaking perturbation DMI that influences  the magnetic critical behavior differently on either side of T$_C$ in the critical regime.

The details of synthesis of the above three single crystals and structural characterization are reported in supplementary material (SM; section 1, Figure S1, Table S1), and the reference therein. Magnetic measurements were carried out on single crystals using a superconducting quantum interference device magnetometer (Quantum Design SQUID‑VSM). For the critical scaling analysis, field-dependent magnetization M(H) measurements were conducted at various temperatures in the magnetic transition region [Figure S3, S7 and S10]. Demagnetization effects are found to be negligible and hence, no correction was applied.

\begin{figure*}
    \centering
    \includegraphics[scale=0.28]{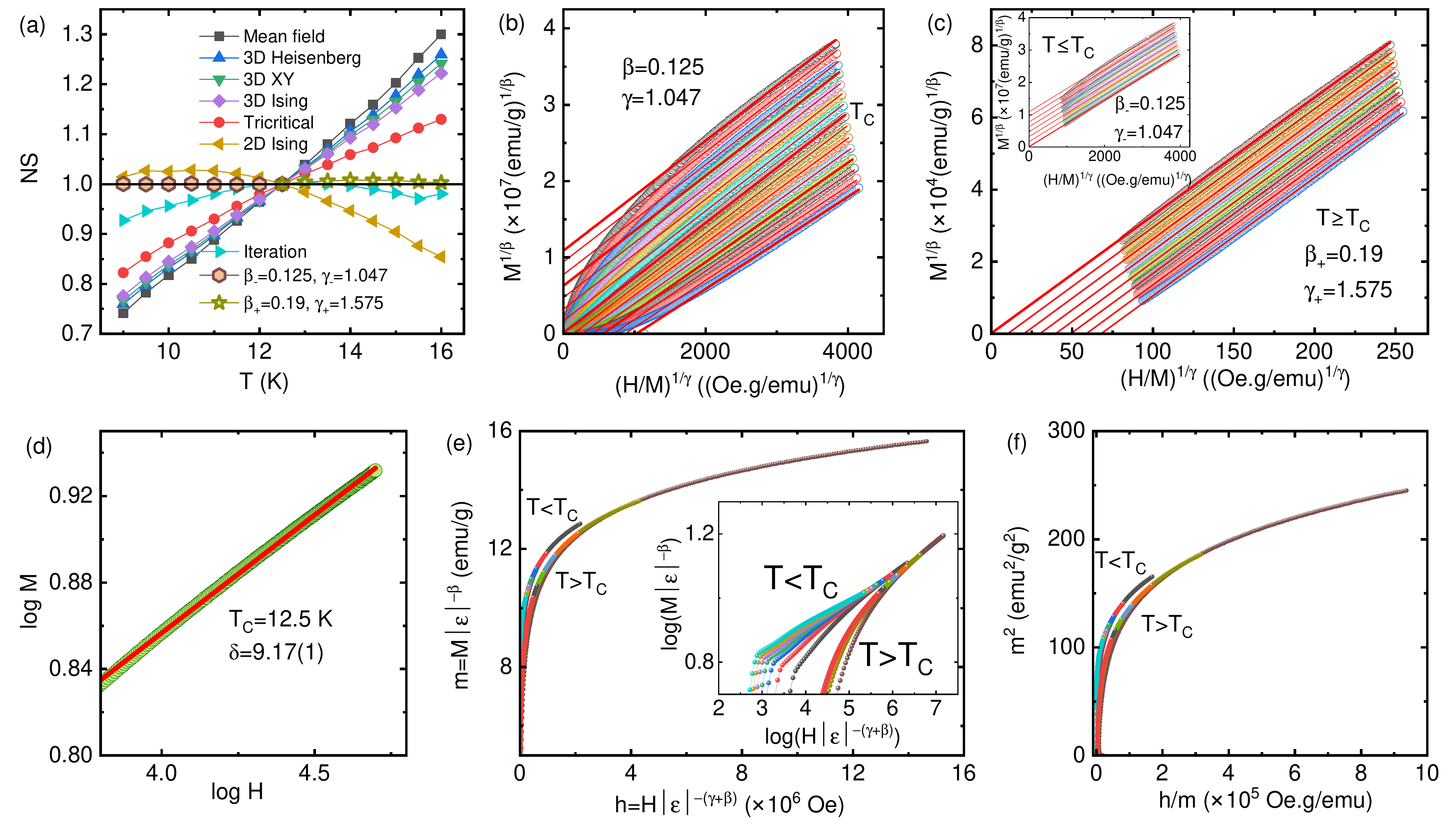}
    \caption{NCS-Br: (a) Normalised slope NS(T) obtained from exponents of various universality classes, iteration method and distinct set of critical exponents $\beta$ and $\gamma$. (b) Linear fitting to the modified Arrott plots (MAPs) using $\beta$=0.125, $\gamma$=1.047 (c) MAPs using $\beta$=0.19, $\gamma$=1.575 for T$\geq$T$_C$. Inset: MAPs using $\beta$=0.125, $\gamma$=1.047 for T$\leq$T$_C$ (d) Log-log plot of magnetic isotherm M(H) at T$_C$=12.5K with a linear fitting (e) Renormalised magnetization ($m$) vs. field ($h$) plot below and above T$_C$. Inset: $m$ vs. $h$ in log-log scale. (f) $m^2$ vs. $h/m$ plots (Renormalised Arrott plots).}
    \label{Criticality-NCS Br}
\end{figure*}
Among the three  quasi-2D OIHPs, (C$_7$H$_9$NBr)$_2$CuX$_4$ (X = Br and Cl) crystallise in noncentrosymmetric monoclinic structure with polar space group \textit{Cc}~\cite{6rgd-57pc}, while (C$_7$H$_{10}$N)$_2$CuBr$_4$ adopt centrosymmetric structure with space group \textit{C2/c} [Figure S1 and Table S1 in SM]. All the compounds show dominant ferromagnetic interaction, verified from the positive Curie-Weiss temperature ($T_{CW}$) [Insets of Figure S2(a),S6(a),S9(a)]. Note that, while both the Br analogs ((C$_7$H$_9$NBr)$_2$CuBr$_4$ (NCS-Br) and (C$_7$H$_{10}$N)$_2$CuBr$_4$ (CS-Br)) exhibit easy saturation along the out-of-plane direction [Figure S2 and S9], (C$_7$H$_9$NBr)$_2$CuCl$_4$ (NCS-Cl) show an in-plane easy saturation [Figure S6]. The criticality analysis is carried out along the easy saturation directions for all samples.

Presence of second-order magnetic phase transition is confirmed from the concave-downward curvature of the Arrott plots (mean field model) [Figure S4(a), S8(a), S11(a)], following the Banerjee criterion~\cite{BANERJEE196416}. The associated critical exponents $\beta$, $\gamma$ and $\delta$ were extracted by fitting spontaneous magnetization (M$_S$) below T$_C$, inverse susceptibility ($\chi_0^{-1}$) above T$_C$, and critical isotherm (M(H) at T$_C$), respectively~\cite{PhysRevB.96.054406}, according to the following relations:
\begin{equation}
    M_S(T)=M_0(-\epsilon)^{-\beta}; \epsilon<0, T<T_C
    \label{eq 1}
\end{equation}
\begin{equation}
    \chi_0^{-1}(T)=(\frac{h_0}{m_0})\epsilon^\gamma; \epsilon>0, T>T_C
    \label{eq 2}
\end{equation}

\begin{equation}
    M=DH^{\frac{1}{\delta}}; \epsilon=0, T=T_C
\label{eq 3}
\end{equation}

where $\epsilon=\frac{T-T_C}{T_C}$ and $M_0$, $\frac{h_0}{m_0}$, $D$ are the critical amplitudes. To study the critical behavior, modified Arrott plots (MAPs) M$^{\frac{1}{\beta}}$ vs. (H/M)$^{1/\gamma}$ are constructed based on the Arrott Noake's relation; $(H/M)^{1/\gamma}=a\epsilon+bM^{\frac{1}{\beta}}$ where \textit{a} and \textit{b} are constants~\cite{PhysRevLett.19.786,PhysRevMaterials.8.114418,px6g-hprm}. The $\beta$ and $\gamma$ are so chosen that the MAPs produce a set of parallel straight lines, and the line at T$_C$ passes through the origin~\cite{PhysRevLett.128.015703}. Further the critical exponents should self consistently satisfy the Widom scaling relation $\delta=1+\frac{\gamma}{\beta}$ ~\cite{PhysRevB.96.134410} where $\delta$ is determined by fitting the critical M(H) isotherm at T$_C$ with the equation~\ref{eq 3}. The parallelism of the MAPs is quantified by evaluating the normalised slope (NS =S(T)/S(T$_C$)), where S(T) and S(T$_C$) are the slope of the linear fit to the MAPs at temperature T and T$_C$ respectively. The accurate critical exponents are those which produce the NS value close to 1 and generally, iteration method is employed (see section 2 of SM) to determine the critical exponents $\beta$ and $\gamma$.

In the critical asymptotic region, the magnetic scaling equation is expressed as 
\begin{equation}
    M(H,\epsilon)=\epsilon^{\beta}f\pm(\frac{H}{\epsilon^{\beta+\gamma}})
    \label{eq 4}
\end{equation}

where f$_+$ and f$_-$ are regular functions correspond to T$>$T$_C$ and T$<$T$_C$ respectively. This equation can also be rewritten as 
\begin{equation}
    m=f\pm(h)
    \label{eq 5}
\end{equation}

where $m=\epsilon^{-\beta}M(H,\epsilon)$ is the renormalized magnetization and $h=\epsilon^{-(\beta+\gamma)}H$ is the renormalized field~\cite{PhysRevB.101.214413}. The equation~\ref{eq 5} implies that for the correct critical exponent $\beta$ and $\gamma$ values, the renormalized $m$ vs. $h$ curves should collapse onto two different universal branches, one below T$_C$ and another above T$_C$~\cite{PhysRevB.96.054406}.

First, we focus on the NCS-Br compound which is reported to possess a transverse conical spiral magnetic ground state driven by the interplay of ferromagnetic exchange, magnetocrystalline anisotropy, and DMI ~\cite{6rgd-57pc}. This makes it an interesting system to examine the underlying magnetic UC in the presence of such competing interactions by analyzing the critical behavior. In this regard, MAPs are generated for all the six UCs [Figure S4 in SM], following the procedure discussed above. The MAPs show better linearity in the high-field region and NS is determined for all UCs from linear fits to the high field region [FIG. ~\ref{Criticality-NCS Br}(a)]. The NS plot shows a large deviation from the constant slope NS=1 line for the 3D UCs and mean-field model, suggesting that the criticality in the compound is unlikely to be governed by 3D critical exponents. Further, while the 2D Ising model exhibits the smallest deviation in NS below T$_C$, the corresponding NS curve deviates more from the constant-slope NS=1 line above T$_C$. In contrast, the tricritical mean field model produces the least deviation above T$_C$. This primarily indicates that the critical exponents governing the critical behavior below and above the T$_C$ could be different. Note that, though the least deviated NS is produced by the tricritical mean field model above T$_C$, the deviation is still very high compared to the NS produced by 2D Ising model below T$_C$.

Subsequently, the iteration method is followed to evaluate the critical exponents. In this method, the initial exponents are chosen from that UC which generates the least deviated slope. Hence 2D Ising model is initially adopted. It is found that, for 2D Ising model, the MAP close to T$_C$ does not pass through the origin. Therefore, keeping $\beta$ (=0.125) fixed, the $\gamma$ value is varied in such a way that the MAP close to T$_C$ passes through the origin. Then iteration method is employed which yields $\beta$=0.147(6) and $\gamma$=0.999(25) as shown in Figure S5 of SM. However, the resulting NS from the iteration method is very much deviated from the NS=1 line, especially below T$_C$ [FIG.~\ref{Criticality-NCS Br}(a)]. Hence, the exponents obtained from iteration are not accurate and the iteration method is not suitable to determine the precise values of critical exponents due to the fact that single set of critical exponents can not capture the critical behavior in the present scenario. This is evident from the MAPs constructed from one set of exponents $\beta$=0.125 and $\gamma$=1.04 that show nearly parallel lines below T$_C$, however a significant deviation from parallelism is observed above T$_C$ [FIG.~\ref{Criticality-NCS Br}(b)].

To circumvent this discrepancy, a new iterative method, proposed by Ghosh \textit{et.al}~\cite{PhysRevLett.128.015703}, is adopted. In this method, first the two UCs which produce the least deviated NS below and above T$_C$ are chosen and then the critical exponents are separately iterated below and above T$_C$, untill getting NS values near to 1. By varying $\beta$ and $\gamma$ around the 2D Ising exponents for T$\leq$ T$_C$ and tricritical mean field exponents for T$\geq$T$_C$, we obtained a distinct set of $\beta$ and $\gamma$, below and above T$_C$ ($\beta_-$ and $\gamma_-$ for T$\leq$T$_C$ and $\beta_+$ and $\gamma_+$ for T$\geq$T$_C$), yielding nearly parallel fitted MAPs with NS(T) values closest to 1 [FIG.~\ref{Criticality-NCS Br}(a)] and the line at T$_C$ passing through the origin [FIG.~\ref{Criticality-NCS Br}(c)]. The optimized exponents are $\beta_-$=0.125, $\gamma_-$=1.047 and $\beta_+$=0.19, $\gamma_+$=1.575, [FIG.~\ref{Criticality-NCS Br}(c)] that leads to $\delta_-$=9.376 and $\delta_+$=9.289 according to the Widom scaling relation $\delta=1+\frac{\gamma}{\beta}$. The value of $\delta$ is further experimentally determined to be 9.17(1) from the linear fits to log M vs. log H plot at T$_C$=12.5 K, following equation ~\ref{eq 3} [FIG.~\ref{Criticality-NCS Br}(d)], indicating that the experimental $\delta$ value is closely similar to the $\delta_-$ and $\delta_+$. FIG.~\ref{Criticality-NCS Br}(e) shows the $m$ vs. $h$ plot with the inset showing the same in the log-log scale by using $\beta_-$=0.125, $\gamma_-$=1.047 for T$<$T$_C$ and $\beta_+$=0.19, $\gamma_+$=1.575 for T$>$T$_C$. Here, all data collapse into two different well divided branches below and above the T$_C$. The deviation from the universal branches in the low-field region observed in the log-log plot [Inset FIG.~\ref{Criticality-NCS Br}(e)] can be correlated to the observed nonlinearity of the MAPs in the low-field region [FIG.~\ref{Criticality-NCS Br}(b)]. The reliability of these critical exponents and T$_C$ have been further corroborated from the renormalized Arott plots $m^2$ vs. $h/m$ as shown in [FIG.~\ref{Criticality-NCS Br}(f)] where all the data also divided into two well separated branches, indicating the consistency of the scaling analysis.

The obtained critical exponents ($\beta$=0.125, $\gamma$=1.047 below T$_C$ and $\beta$=0.19, $\gamma$=1.575 above T$_C$) lie within the range associated with 2D XY/h$_4$ UC ($\beta$=0.125-0.23) where $\beta$=0.125 corresponds to 2D Ising and $\beta$=0.23 to 2D XY limit~\cite{doi:10.1126/science.1221878,doi:10.1126/science.abd5146,Taroni_2008}. However, it is worth noting that $\beta$=0.19 has been identified to represent 2D XY UC in previous studies~\cite{PhysRevLett.64.32, PhysRevB.47.8461, Bramwell1993}. Thus, the change in $\beta$= 0.125 below T$_C$ to $\beta$=0.19 above T$_C$, indicates a change in UC from 2D Ising to 2D XY-like across the magnetic transition in NCS-Br. Such unconventional critical behavior has been attributed earlier to the interplay of DMI, symmetric exchange, and magnetocrystalline anisotropy in Cu$_2$OSeO$_3$~\cite{PhysRevLett.128.015703}. For NCS-Br, our recent magnetic study indicates the possible existence of a non-collinear conical spiral magnetic ground state induced by the DMI~\cite{6rgd-57pc}. We therefore propose that the observed unconventional criticality may be linked to the underlying DMI. To establish whether DMI underlies this anomalous criticality in NCS-Br, we have undertaken detailed critical scaling analysis for the isostructural Cl analogue and centrosymmetric Br analogue, where DMI is expected to be relatively weaker.

\begin{figure}
    \centering
    \includegraphics[scale=0.17]{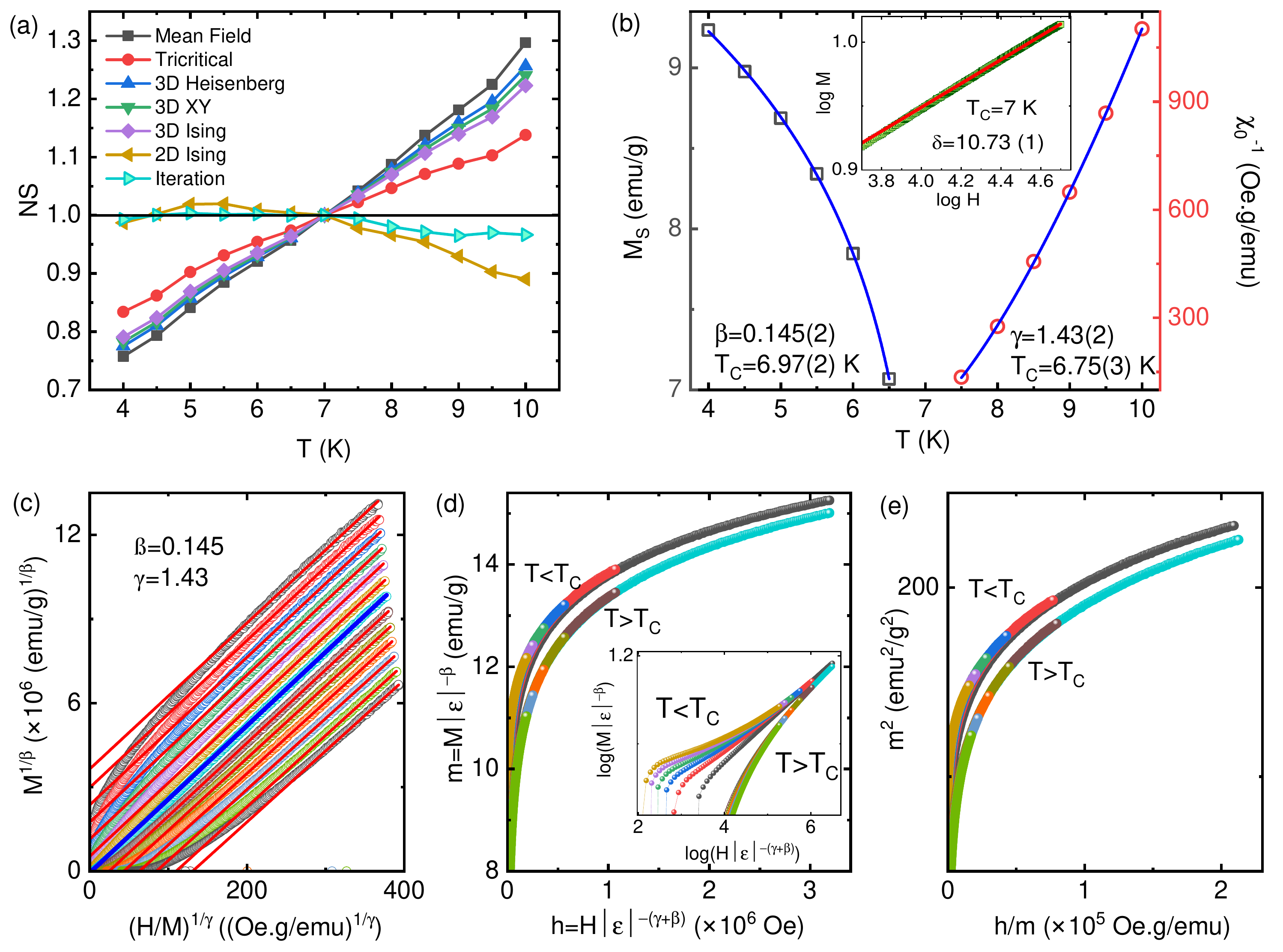}
    \caption{NCS-Cl: (a) Normalised slope NS(T) obtained from exponents of various universality classes and iteration method. (b) Temperature dependent spontaneous magnetization M$_S$(T) (left) and inverse susceptibility $\chi_0^{-1}$ (right) with fitted solid curves (blue curves). Inset: Linear fitting of M vs. H at T$_C$ plot in log-log scale. (c) Modified Arrott plots (MAPs) using $\beta$=0.144, $\gamma$=1.429 (d) Renormalised magnetization ($m$) vs. field ($h$) plot below and above T$_C$. Inset: $m$ vs. $h$ in log-log scale. (e) $m^2$ vs. $h/m$ plots.}
    \label{Criticality_NCS-Cl}
\end{figure}

Now we examine the non-centrosymmetric NCS-Cl compound, which shows a nominally uniaxial in-plane magnetic anisotropy induced by an in-plane DMI vector, as discussed in the earlier work [Figure S6 of~\cite{6rgd-57pc}]. However, the strength of DMI in the NCS-Cl is weaker compared to the NCS-Br compound, consistent with reduced spin-orbit coupling associated with the lighter Cl ligand~\cite{PhysRevMaterials.8.024409, PhysRevB.102.115162,Chang2020}. Upon analyzing the critical behavior, it is found that the NS plot (obtained from the linear fitting of MAPs in the high field region [Figure S8 in SM]) shows the least deviation for 2D Ising UC throughout the transition regime [FIG.~\ref{Criticality_NCS-Cl}(a)]. Iteration method produces the least deviated NS [FIG.~\ref{Criticality_NCS-Cl}(a)] with $\beta$=0.145(2), T$_C$=6.97(2) K from the fitting of M$_S$(T) and $\gamma$=1.43(2), T$_C$=6.75(3) K from the fitting of $\chi_0^{-1}$(T) [FIG.~\ref{Criticality_NCS-Cl}(b)]. Considering this $\beta$ and $\gamma$, the MAP produces a set of straight lines in the high field region with the line at 7 K (T$_C$), passing through the origin [FIG.~\ref{Criticality_NCS-Cl}(c)]. The linear fitting to M-H isotherm at T$_C$=7 K in log-log scale yields $\delta$=10.73(1)[Inset of FIG.~\ref{Criticality_NCS-Cl}(b)], indicating that exponents obey Widom scaling relation. Further, the renormalised m vs. h (along with the plots in their log-log scale) [FIG.~\ref{Criticality_NCS-Cl}(d)] and m$^2$ vs. h/m [FIG.~\ref{Criticality_NCS-Cl}(e)] curves collapse into two different branches below and above T$_C$, showing reliability of the critical exponents. The NS curve obtained from iterative method using $\beta$=0.145, $\gamma$=1.43, deviates slightly from the NS=1 line above T$_C$ and remains close to NS=1 line below T$_C$. This weak asymmetry in NS curve possibly arises from the non-negligible DMI present in this compound. The obtained critical exponents ($\beta$=0.144, $\gamma$=1.429 and $\delta$=10.73), are characteristics of the 2D Ising UC which can be correlated to the dominant unidirectional in-plane spin-orientation in the NCS-Cl compound [Figure S6 of~\cite{6rgd-57pc}]. Although Ising universality is often commonly associated with an out-of-plane easy axis, it can also be realized in systems with in-plane magnetization, having uniaxial magnetic anisotropy, as reported in earlier studies for Fe/W(110) and Fe(110)/Ag(111) films~\cite{PhysRevLett.67.1646,Back1995}. Hence, the 2D Ising exponents consistently explain the magnetic response observed in NCS-Cl compound.

\begin{figure}
    \centering
    \includegraphics[scale=0.175]{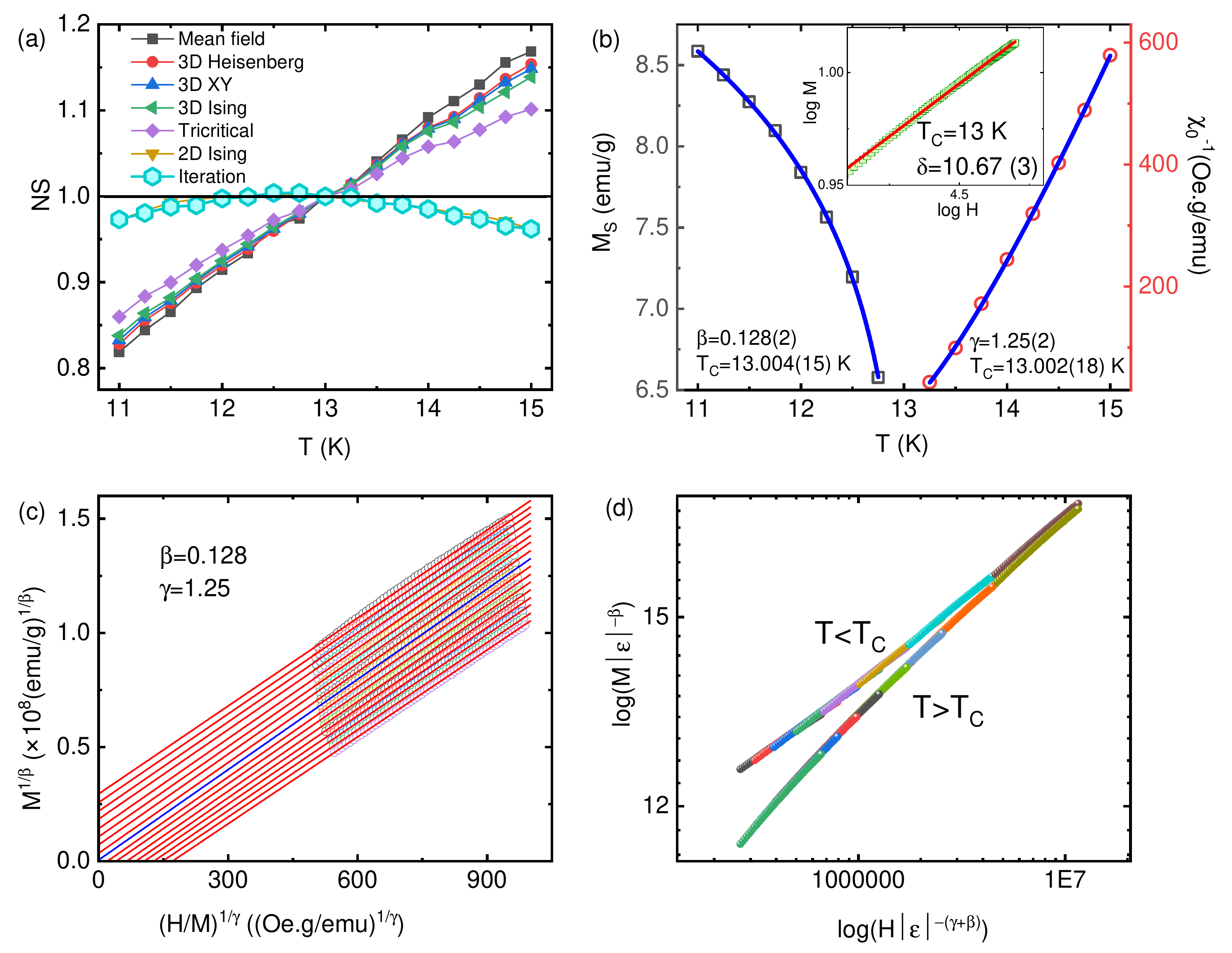}
    \caption{CS-Br: (a) Normalised slope NS(T) obtained from exponents of various universality classes and iteration method. (b) Temperature dependent spontaneous magnetization M$_S$(T) (left) and inverse susceptibility $\chi_0^{-1}$ (right) with fitted solid curves (blue curves). Inset: Linear fitting of M vs. H at T$_C$ plot in log-log scale. (c) Modified Arrott plots (MAPs) using $\beta$=0.128, $\gamma$=1.25 (d) Log-log plot of renormalised magnetization ($m$) vs. field ($h$) below and above T$_C$ in the high field region.}
    \label{Criticality_CS-Br}
\end{figure}

Finally, we delve into the critical behavior of the centrosymmetric compound CS-Br, showing an out-of-plane easy axis. Here, 2D Ising UC produces NS close to 1 both below and above T$_C$, [FIG.~\ref{Criticality_CS-Br}(a) and Figure S11], indicating the presence of conventional critical behavior. Further, the iteration method produces the least deviated NS [FIG.~\ref{Criticality_CS-Br}(a)] with $\beta$=0.128(2), T$_C$=13.004(15) K from the fitting of M$_S$(T) and $\gamma$=1.25(2), T$_C$=13.002(18) K from the fitting of $\chi_0^{-1}$(T)[FIG.~\ref{Criticality_CS-Br}(b)]. The exponents $\beta$=0.128 and $\gamma$=1.25 produce a set of straight lines in MAPs with the line at T$_C$$\sim$ 13 K passing through origin [FIG.~\ref{Criticality_CS-Br}(c)] and the exponents obey Widom scaling relation as concluded from the $\delta$ value ($\sim$10.67(3)) shown in the inset of FIG.~\ref{Criticality_CS-Br}(b). Further, the log-log plot of m vs. h curves produces two distinct branches below and above T$_C$ in the high field region [FIG.~\ref{Criticality_CS-Br}(d)], which suggests the accuracy of the critical exponents. The value of $\beta$=0.128 indicates the presence 2D Ising-type spin interaction, consistent with the out-of-plane easy axis observed in CS-Br compound.

\begin{figure}
    \centering
    \includegraphics[width=0.9\linewidth]{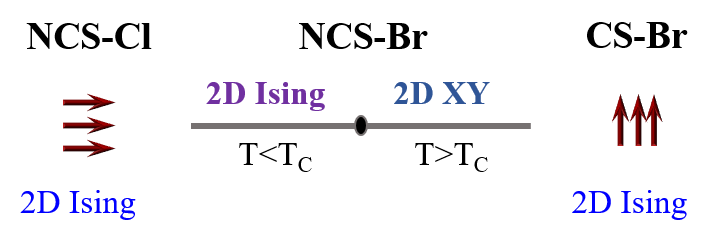}
    \caption{Spectrum of critical behavior}
    \label{last figure}
\end{figure}

 The UCs obtained from the critical scaling analysis has been summarized in FIG.~\ref{last figure}, showing the spectrum of critical phenomena in the hybrid perovskites caused by changing the crystal symmetry and halide ligands. In all three compounds, the $\beta$ values deviate significantly from 3D UCs  with $\beta$$\sim$0.32 to 0.37 and lie in the range of 0.125-0.23, corresponding to 2D XY/h$_4$ UC~\cite{PhysRevB.110.L060405, doi:10.1126/science.1221878}. In 2D XY/h$_4$ UC, $\beta$=0.125 represents the extreme anisotropic 2D Ising limit while $\beta$=0.23 signifies the isotropic 2D XY limit. Hence, it can be unambiguously concluded that the magnetic criticality in these compounds is two dimensional. However, the $\delta$ values for all three compounds ($\approx$10) lie in between the 3D ($\delta$$\approx$5) and 2D UCs ($\delta$$\approx$15)~\cite{PhysRevB.110.L060405,Kosterlitz1974,hirakawa1982kosterlitz}, indicating the presence of non-negligible interlayer magnetic coupling.

The asymmetric critical behavior observed in NCS-Br is particularly intriguing. As discussed in the introduction, asymmetric 3D critical exponents on either side of T$_C$ have been reported in Cu$_2$OSeO$_3$~\cite{PhysRevLett.128.015703}, Cr$_{\frac{1}{3}}$TaS$_2$~\cite{PhysRevB.107.144425}, MnBi$_4$Te$_7$ ~\cite{y5tx-5vm7} and FePd$_2$Te$_2$~\cite{10.1063/5.0300224}. However, the present study highlights the asymmetric criticality in 2D regime, where 2D Ising-type criticality below T$_C$ evolves into 2D XY-type critical behavior above T$_C$ in NCS-Br. In contrast, investigation on iso-structural Cl and centrosymmetric Br analogues reveals conventional critical behavior. This highlights the role of DMI, which is more effective in NCS-Br. We conjecture that DMI influences the critical fluctuations and correlations differently on the two sides of the transition, thereby favoring distinct effective critical regimes. Interestingly, the difference between low-T and high-T 2D exponents is found to be larger in NCS-Br than that observed in most of the 3D cases. This suggests that reduced dimensionality may provide a more favorable setting to explore the effect of dangerously irrelevant perturbation, yielding different critical behavior on the two sides of T$_C$. Our results motivate theoretical and numerical studies of asymmetric criticality in quasi-2D magnets, where controlled 2D simulations may provide direct tests of the distinct critical behavior above and below T$_C$ and the role of DMI.

In summary, we have demonstrated the presence of asymmetric critical exponents on two sides of the magnetic transition in 2D regime for a DMI-active NCS-Br compound. Notably, it exhibits 2D Ising and 2D XY type criticality below and above T$_C$, respectively. In contrast, investigation on
additional two hybrid perovskites NCS-Cl and CS-Br, for which the DMI contribution is expected to be less significant, reveals conventional 2D critical behavior. The systematic contrast in the critical behavior among these compounds points to DMI as a key symmetry-breaking perturbation for the observed asymmetric criticality in NCS-Br compound. 

\textit{Acknowledgments}- P.B. acknowledges Prof. B. Delamotte, Prof. Stuart S. P. Parkin and Prof. S.N. Kaul for valuable insights. P.B. thanks Dr. Amitava Ghosh and Dr. Hanuma Kumar Dara for useful discussions.

\nocite{*}

\bibliography{apssamp}

\normalsize\clearpage
\begin{onecolumngrid}
	\begin{center}
		{\fontsize{12}{12}\selectfont
			\textbf{Supplementary Material: Different critical exponents on two sides of the magnetic transition in two dimensions\\}}

{\normalsize P. Biswal\textsuperscript{1,2,*}, Ruma Khatun\textsuperscript{1,2}, G. Tripathy\textsuperscript{1,2}, Diptikanta Swain\textsuperscript{3}, D. Samal\textsuperscript{1,2,$\dagger$}\\
\textsuperscript{1}Institute of Physics, Bhubaneswar 751005, India\\
\textsuperscript{2}Homi Bhabha National Institute, Anushaktinagar, Mumbai 400094, India\\\textsuperscript{3}Institute of Chemical Technology–IndianOil Odisha Campus, Bhubaneswar 751013, India}
	\end{center}

\newcounter{defcounter}
	\setcounter{defcounter}{0}
	\setcounter{figure}{0}
  \renewcommand{\figurename}{Figure}
	\renewcommand{\thefigure}{S\arabic{figure}}
	\renewcommand{\tablename}{Table}
    \renewcommand{\thetable}{S\arabic{table}}
\renewcommand{\thesection}{\arabic{section}}
\setcounter{section}{0}

\vspace{0.03cm}	

\section{Synthesis and structural characterization}
\subsection{Single crystal synthesis and structural characterization of NCS-Br and NCS-Cl}
The details of the synthesis and structural characterization of NCS-Br and NCS-Cl compound are reported in a recent paper~\cite{6rgd-57pc}.

\subsection{Single crystal synthesis and structural characterization of CS-Br}
CS-Br ((C$_7$H$_{10}$N)$_2$CuBr$_4$) 
single crystals were synthesized from solution evaporation method. First, benzylammonium salt was synthesized from the homogeneous solution of benzylamine and HBr. Then this salt and CuBr$_2$ were added in a stoichiometric ratio and a homogeneous solution was formed by further adding HBr and distilled water and stirring at 80$^o$ C. Then the solution was kept inside an oven at 80$^o$ C. Single crystals were obtained after 7 days.

The crystal structure and structural information are presented in Figure~\ref{Crystal structure of CS-Br} and Table~\ref{single crystal data of CS-Br} respectively.

\section{Iteration Method}
The method involves initially selecting the critical exponents $\beta$ and $\gamma$ of that UC for which NS plot shows the least deviation from 1. Using the corresponding $\beta$ and $\gamma$ of that particular UC, MAPs are constructed. Subsequently, the M$_S$ and $\chi_0^{-1}$ values are obtained from the y and x-axis intercepts of the linear fitting to the MAPs respectively. From the fitting of M$_S$(T) and $\chi_0^{-1}$(T) using equation 1 and equation 2 (mentioned in the main text) respectively, a new set of $\beta$ and $\gamma$ is obtained. Then MAPs are reconstructed using the new set of $\beta$ and $\gamma$. The above procedure is iterated until getting stable values of $\beta$ and $\gamma$. The stable $\beta$ and $\gamma$ obtained from this iteration method are independent of the initial parameters which denotes that the critical exponents obtained from the iteration method are intrinsic to the system.

\begin{figure}[H]
    \centering
   \includegraphics[scale=0.7]{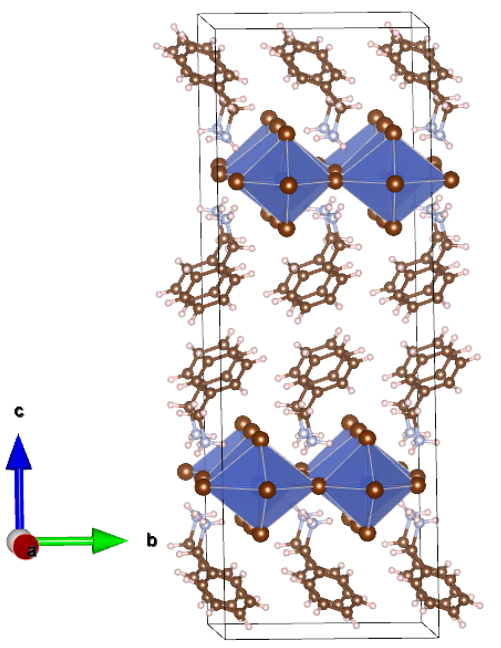}
    \caption{Crystal structure of CS-Br ((C$_7$H$_{10}$N)$_2$CuBr$_4$)}
    \label{Crystal structure of CS-Br}
\end{figure}

\begin{table}[H]
\setlength{\tabcolsep}{0.11 cm}
\caption{Structural information of CS-Br obtained from the single crystal XRD measurements at room temperature.}
    \centering
    \begin{tabular}{cccc}
    \hline \hline
    \textbf{Parameters}&\textbf{CS-Br}\\
  \hline
         Empirical formula&(C$_7$H$_{10}$N)$_2$CuBr$_4$\\ 
         Formula weight (g/mol)&599.5\\
         Temperature (K)&296
\\Wavelength (\AA)&0.71073
\\Crystal system&Monoclinic\\Space group&\textit{C 2/c}\\
         Unit cell dimension&a = 10.953(3)\AA  \\
         &b = 11.026(2)\AA \\
          &c = 31.519(7)\AA\\
          & $\alpha$ = 90$^\circ$ \\ 
          & $\beta$ = 100.034(8)$^\circ$ \\
          & $\gamma$ = 90$^\circ$\\ Volume (\AA$^3$)&3748.4(14)\\Z&8\\Crystal density (g/cm$^3$)&2.125 \\F(000)&2296\\Absorption 
          coefficient& 9.688
\\Reflections  collected&23923\\Goodness of fit&1.223
\\Final R indices (\textit{I}\textgreater2$\sigma$(\textit{I}))&R$_{obs}$=0.1510\\& wR$_{obs}$=0.3831\\R indices (all data)& R$_{all}$=0.1916\\&wR$_{all}$= 0.4127\\
          \hline
    \end{tabular}
    \label{single crystal data of CS-Br}
\end{table}

\begin{figure}[H]
    \centering
   \includegraphics[scale=0.3]{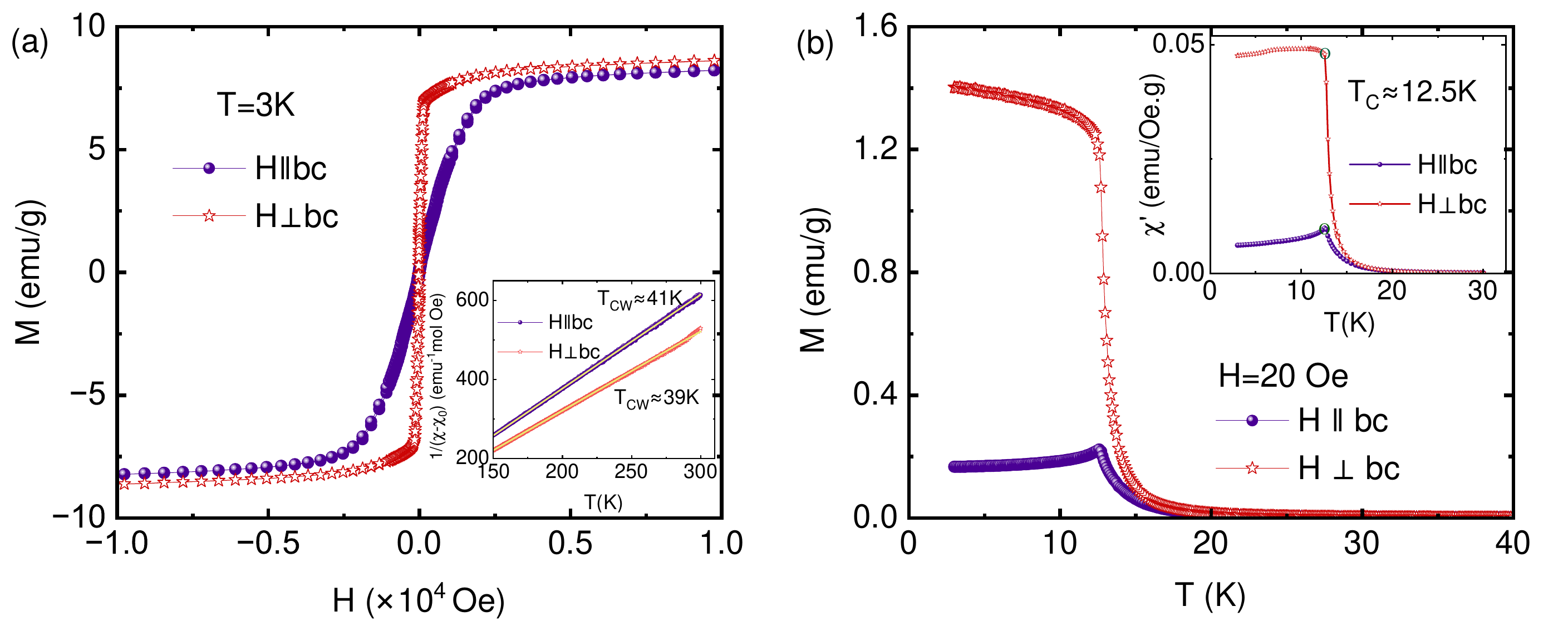}
    \caption{NCS-Br: (a) M-H at 3K along H$\parallel$bc (In-plane orientation) and along H$\perp$bc (Out-of-plane orientation). Inset: Fitting of inverse susceptibility utilizing modified Curie-Weiss law ($ \chi (T)=\chi_0+\frac{C}{T-T_{CW}}$; $\chi_0$ is temperaure independent contribution to the susceptibility, C is the Curie constant and $T_{CW}$ is the Curie-Weiss temperature), showing Curie-Weiss temperature T$_{CW}$$\sim$ 41 K and 39 K along in-plane and out-of-plane direction. (b) M-T at H=20 Oe with inset showing transition temperature T$_C$$\sim$12.5 K in ac susceptibility $\chi$'(T) plot along H$\parallel$bc (In-plane orientation) and along H$\perp$bc (Out-of-plane orientation).}
    \label{magnetic data of S9}
\end{figure}

\begin{figure}[H]
    \centering
   \includegraphics[scale=0.3]{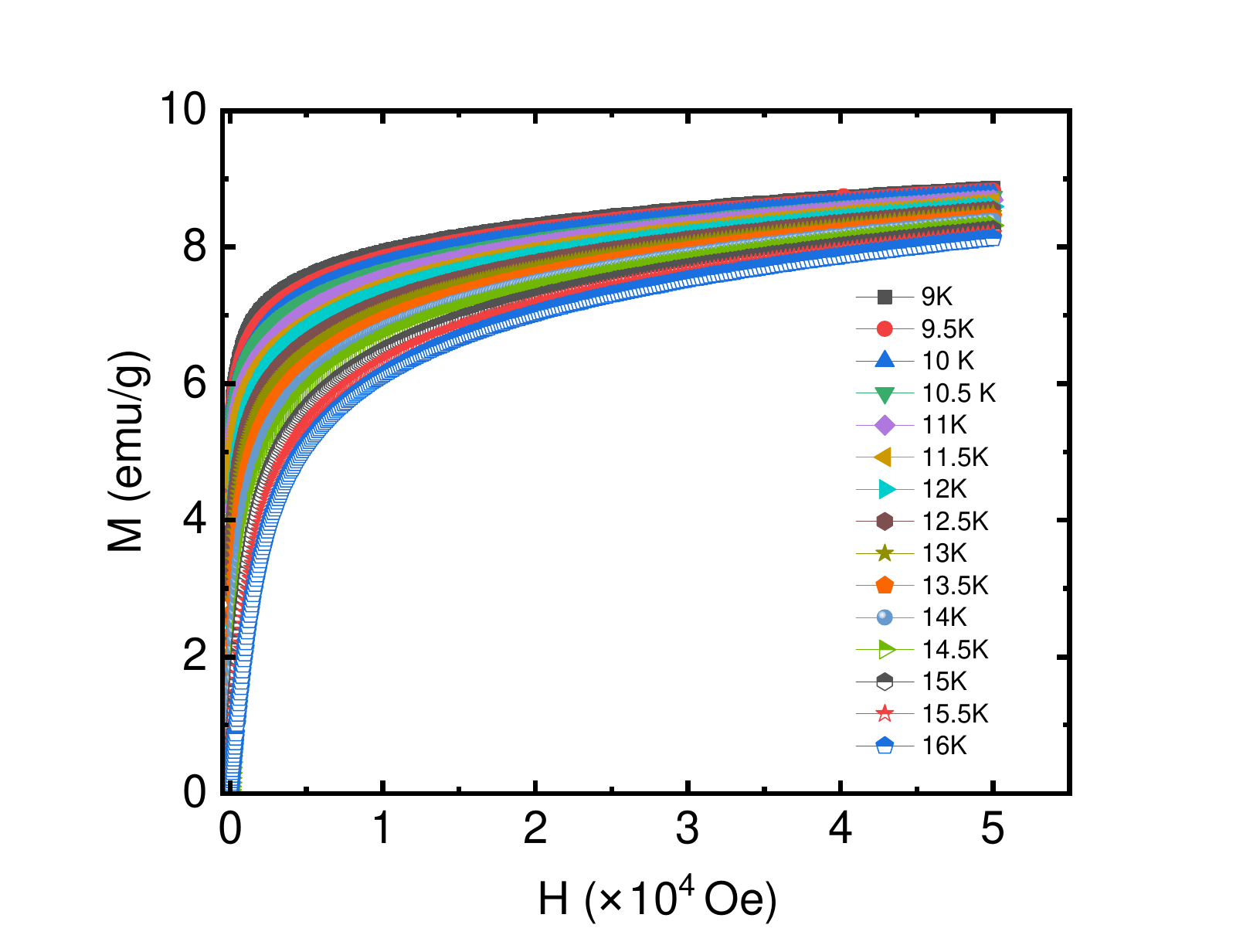}
    \caption{NCS-Br: M-H at various temperatures in the transition regime.}
    \label{S9-MH}
\end{figure}

\begin{figure}[H]
    \centering
   \includegraphics[scale=0.2]{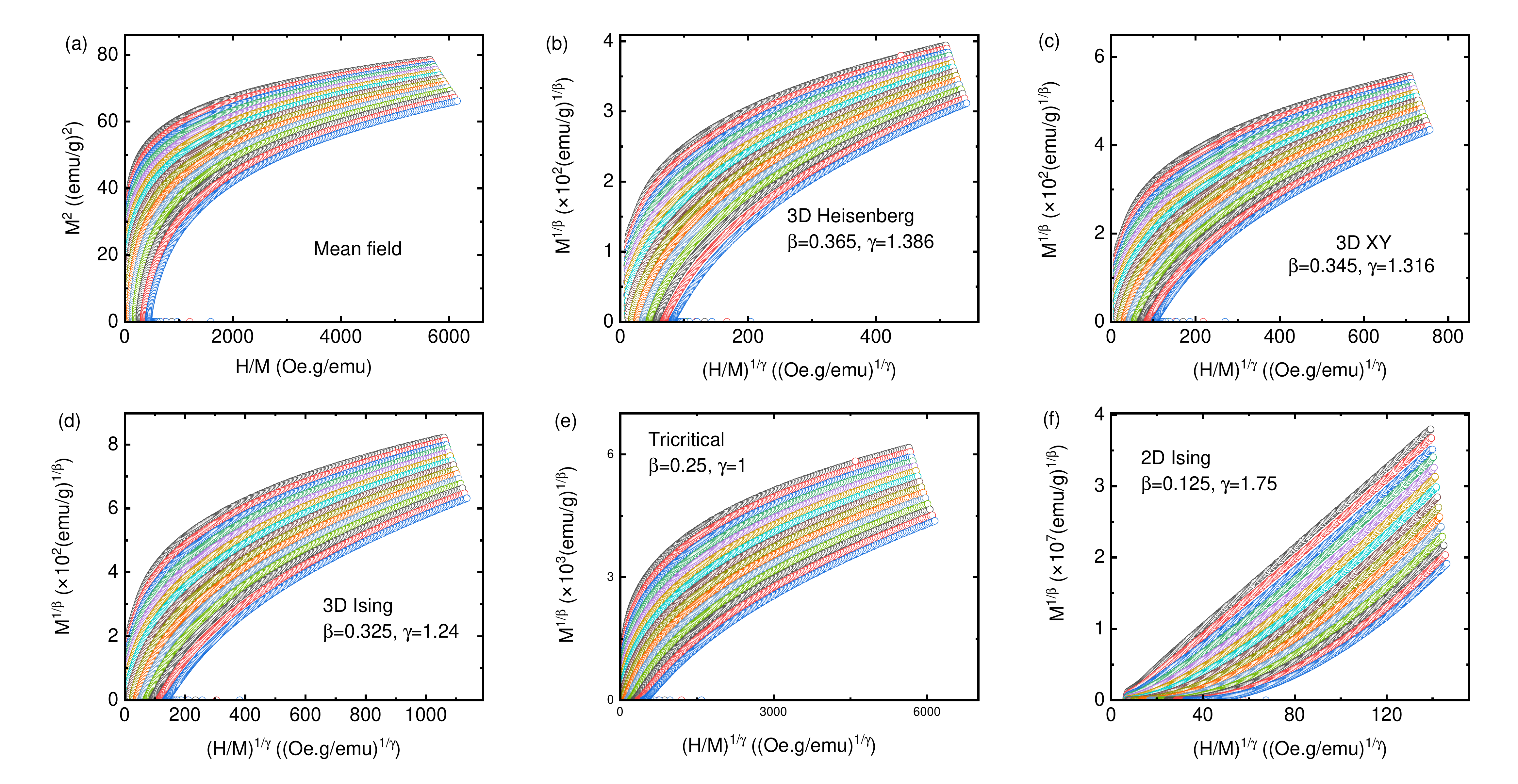}
    \caption{NCS-Br: Modified Arrott plots for various universality classes including mean field model}
    \label{S9-UCs}
\end{figure}
\begin{figure}[H]
    \centering
   \includegraphics[scale=0.3]{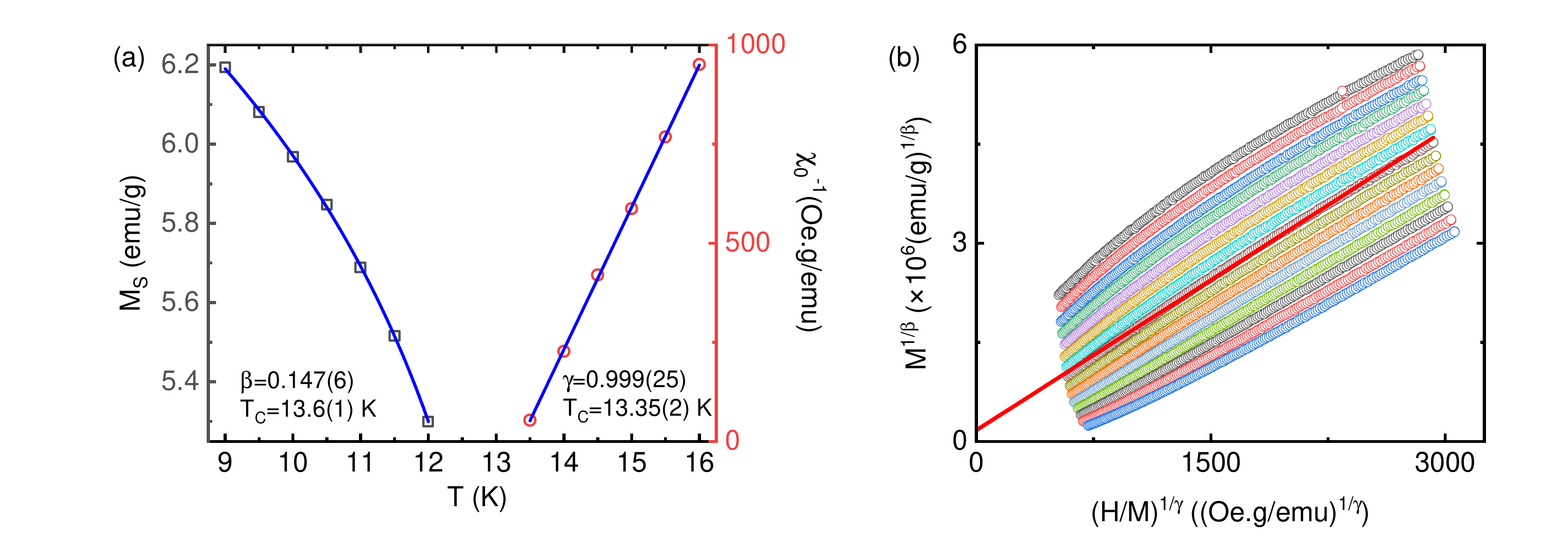}
    \caption{NCS-Br: (a)Temperature dependent spontaneous magnetization M$_S$(T) (left) and inverse susceptibility $\chi_0^{-1}$ (right) with fitted solid curves (blue curves). (b) Modified Arrott plots (MAPs) using $\beta$=0.147, $\gamma$=0.999}
    \label{S9-ITERATION}
\end{figure}
\begin{figure}[H]
    \centering
   \includegraphics[scale=0.3]{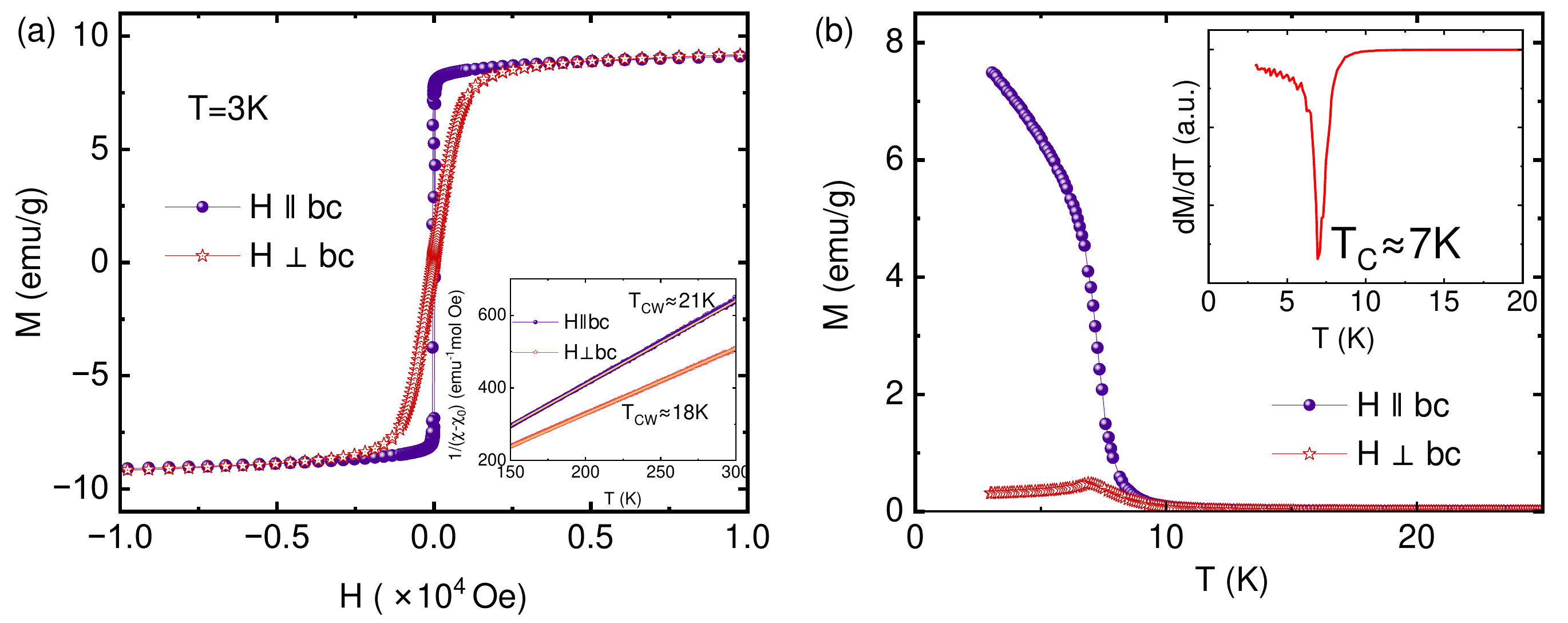}
    \caption{NCS-Cl: (a) M-H at 3K along H$\parallel$bc (In-plane orientation) and along H$\perp$bc (Out-of-plane orientation). Inset: Fitting of inverse susceptibility utilizing modified Curie-Weiss law ($ \chi (T)=\chi_0+\frac{C}{T-T_{CW}}$; $\chi_0$ is temperaure independent contribution to the susceptibility, C is the Curie constant and $T_{CW}$ is the Curie-Weiss temperature), showing Curie-Weiss temperature T$_{CW}$$\sim$ 21 K and 18 K along in-plane and out-of-plane direction. (b) M-T at H=20 Oe along H$\parallel$bc (In-plane orientation) and along H$\perp$bc (Out-of-plane orientation) with inset showing minima of the dM/dT plot as the transition temperature T$_C$$\sim$7 K for 10 Oe applied field along in-plane orientation. The value of T$_C$$\sim$7 K is consistent with the ac susceptibility peak value reported in the recent paper~\cite{6rgd-57pc}.}
    \label{S4-magnetic data}
\end{figure}\begin{figure}[H]
    \centering
   \includegraphics[scale=0.3]{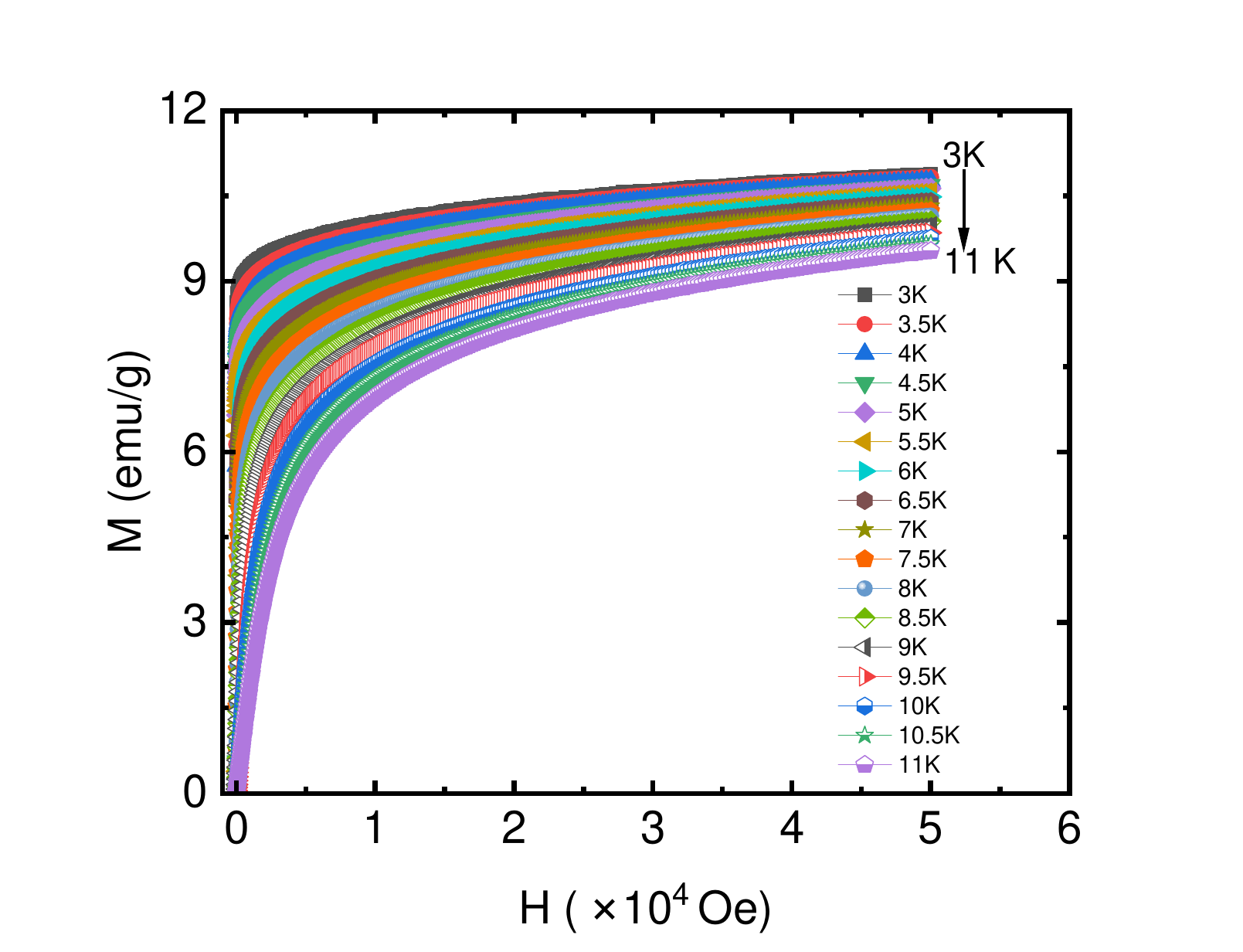}
    \caption{NCS-Cl: M-H at various temperatures in the transition regime.}
    \label{S4-MH}
\end{figure}\begin{figure}[H]
    \centering
   \includegraphics[scale=0.2]{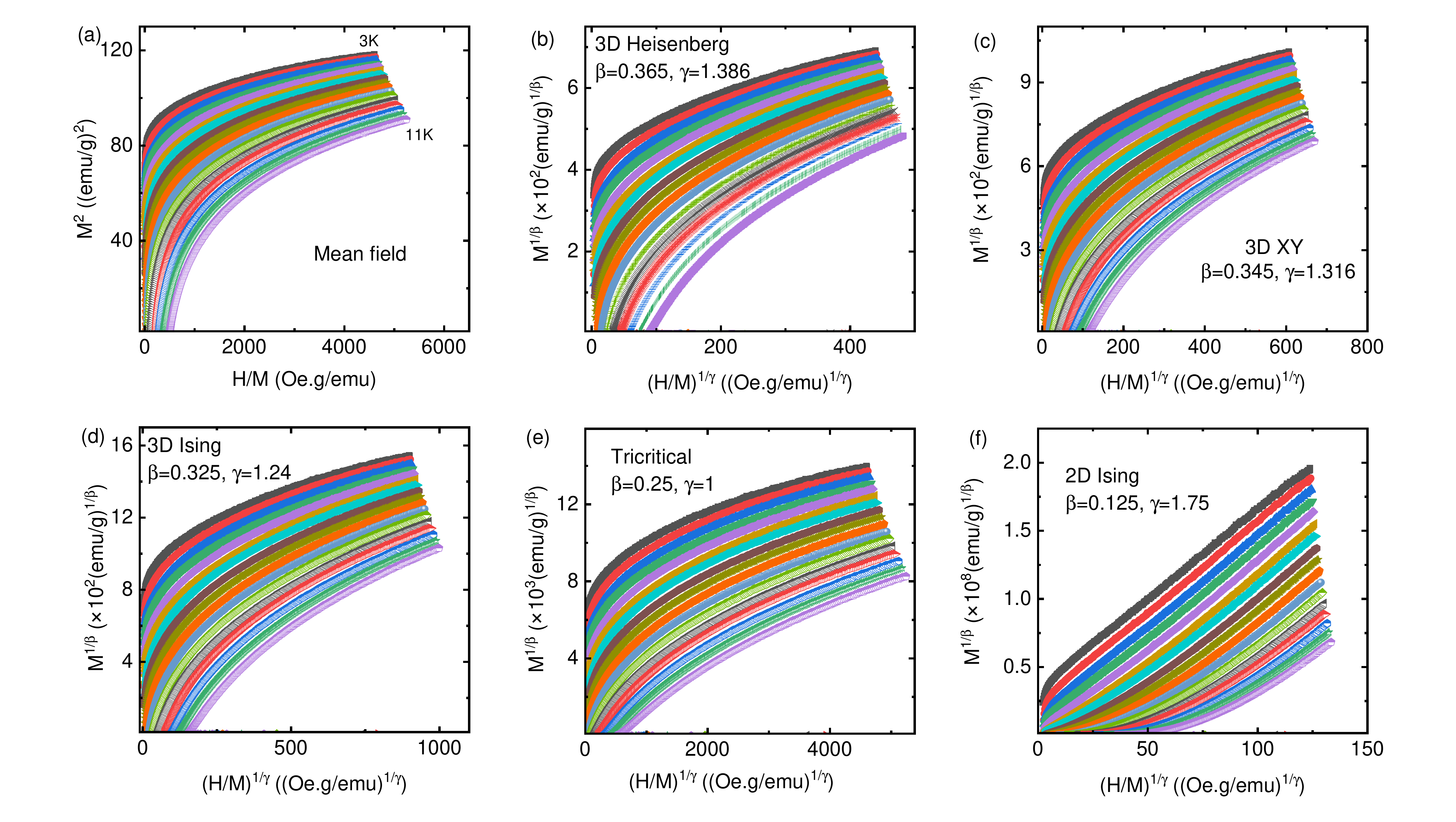}
    \caption{NCS-Cl: Modified Arrott plots for various universality classes including mean field model}
    \label{S4-UCS}
\end{figure}\begin{figure}[H]
    \centering
   \includegraphics[scale=0.3]{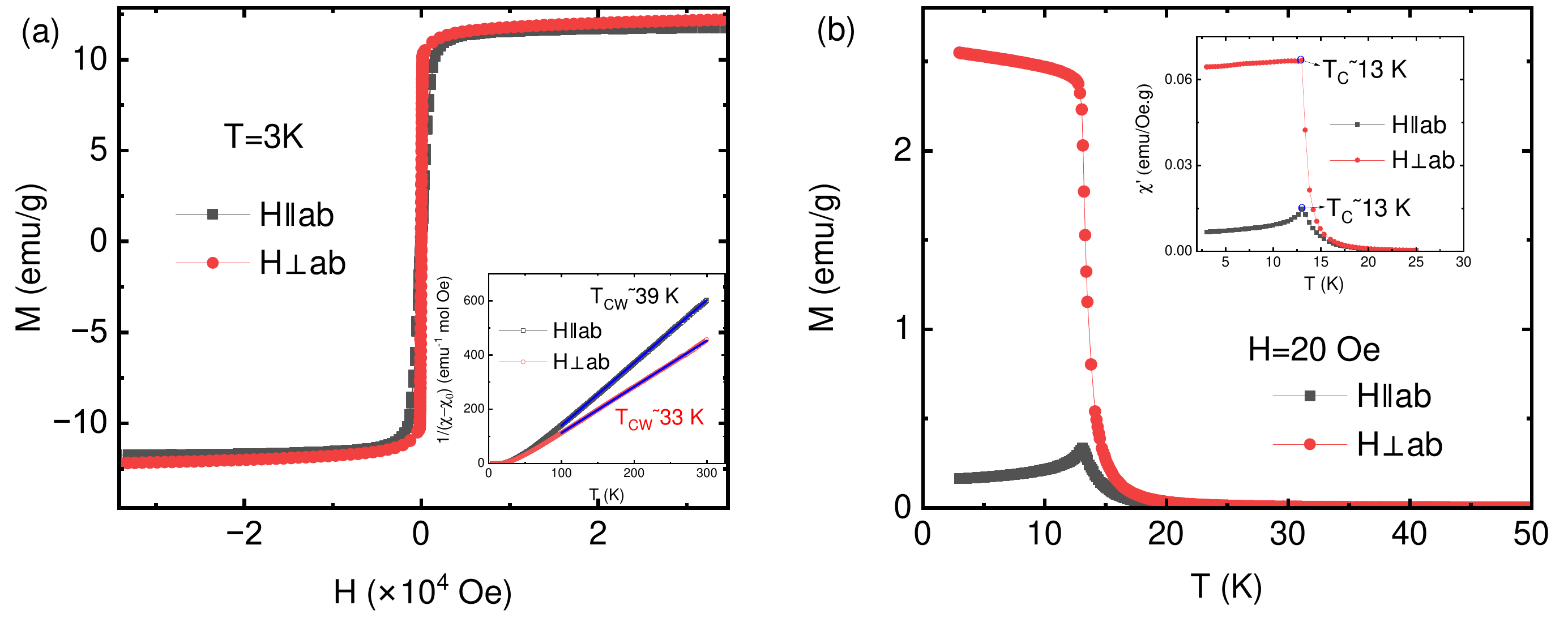}
    \caption{CS-Br: (a) M-H at 3K along H$\parallel$ab (In-plane orientation) and along H$\perp$ab (Out-of-plane orientation). Inset: Fitting of inverse susceptibility utilizing modified Curie-Weiss law ($ \chi (T)=\chi_0+\frac{C}{T-T_{CW}}$; $\chi_0$ is temperaure independent contribution to the susceptibility, C is the Curie constant and $T_{CW}$ is the Curie-Weiss temperature), showing Curie-Weiss temperature T$_{CW}$$\sim$ 39 K and 33 K along in-plane and out-of-plane direction. (b) M-T at H=20 Oe along H$\parallel$ab (In-plane orientation) and along H$\perp$ab (Out-of-plane orientation) with inset showing transition temperature T$_C$$\sim$13 K in ac susceptibility $\chi$'(T) plot along In-plane orientation and along Out-of-plane orientation.}
    \label{S64-magnetic data}
\end{figure}\begin{figure}[H]
    \centering
   \includegraphics[scale=0.3]{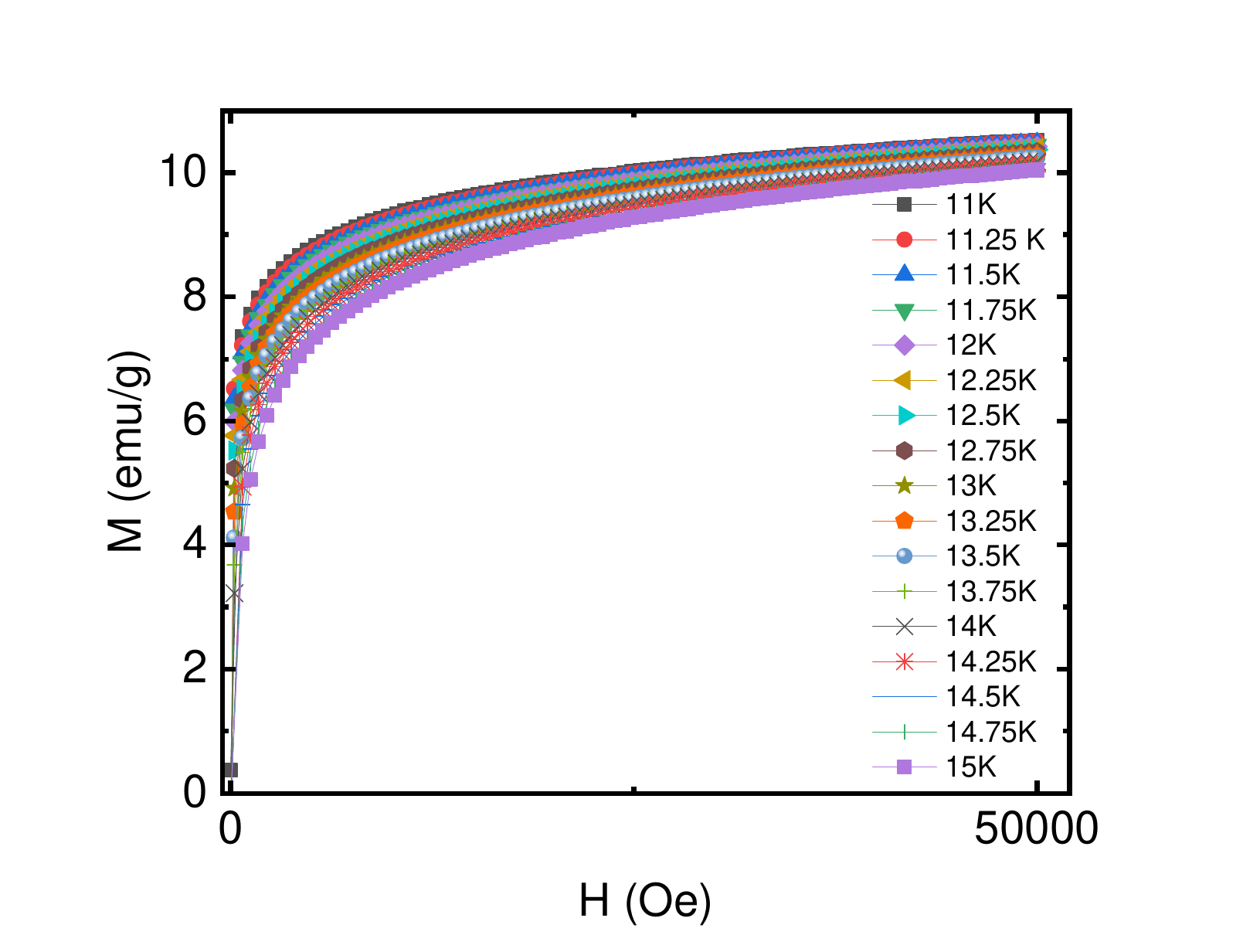}
    \caption{CS-Br: M-H at various temperatures in the transition regime.}
    \label{S64-MH}
\end{figure}
\begin{figure}[H]
    \centering
   \includegraphics[scale=0.2]{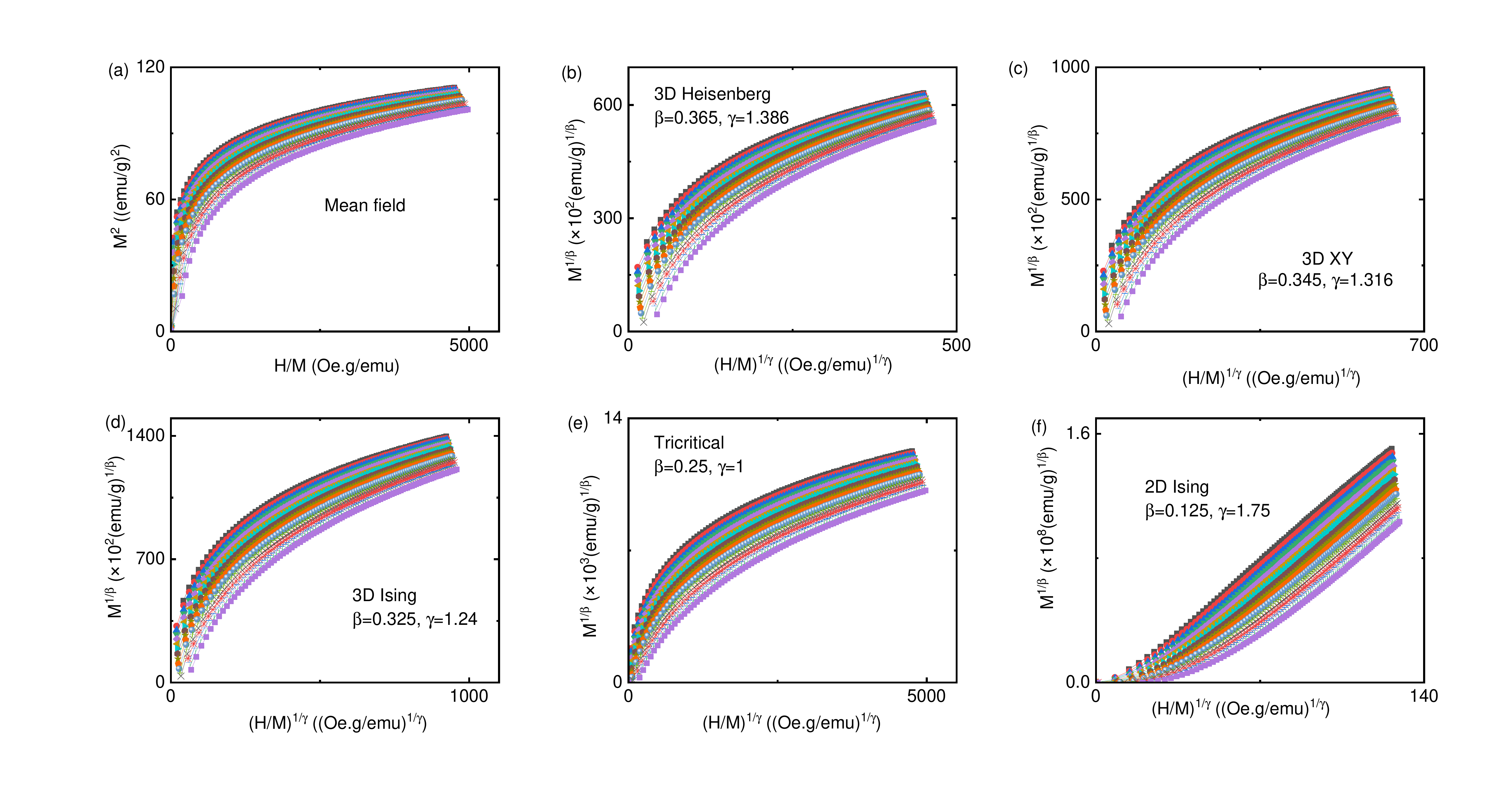}
    \caption{CS-Br: Modified Arrott plots for various universality classes including mean field model}
    \label{S64-UCS}
\end{figure}

\end{onecolumngrid}	

\end{document}